\documentclass[aps,prb,twocolumn]{revtex4-2}
\usepackage[utf8]{inputenc}
\usepackage[T1]{fontenc}

\usepackage{amsmath}
\usepackage{amsfonts}
\usepackage{amssymb}
\usepackage{dsfont}

\usepackage[english]{babel}

\usepackage{physics}

\usepackage{hyphenat}
\usepackage{hyperref}
\usepackage{xcolor}
\usepackage{xspace}
\usepackage{graphicx}

\DeclareMathOperator*{\SumInt}{%
\mathchoice%
  {\ooalign{$\displaystyle\sum$\cr\hidewidth$\displaystyle\int$\hidewidth\cr}}
  {\ooalign{\raisebox{.14\height}{\scalebox{.7}{$\textstyle\sum$}}\cr\hidewidth$\textstyle\int$\hidewidth\cr}}
  {\ooalign{\raisebox{.2\height}{\scalebox{.6}{$\scriptstyle\sum$}}\cr$\scriptstyle\int$\cr}}
  {\ooalign{\raisebox{.2\height}{\scalebox{.6}{$\scriptstyle\sum$}}\cr$\scriptstyle\int$\cr}}
}

\usepackage[normalem]{ulem}

\begin{document}

\title{Melting of devil's staircases in the long-range Dicke-Ising model}
\author{Jan Alexander Koziol}
\affiliation{Department Physik, Staudtstra{\ss}e 7, Universit\"at Erlangen-N\"urnberg, D-91058 Erlangen, Germany}
\author{Anja Langheld}
\affiliation{Department Physik, Staudtstra{\ss}e 7, Universit\"at Erlangen-N\"urnberg, D-91058 Erlangen, Germany}
\author{Kai Phillip Schmidt}
\affiliation{Department Physik, Staudtstra{\ss}e 7, Universit\"at Erlangen-N\"urnberg, D-91058 Erlangen, Germany}

\begin{abstract}
We present ground-state phase diagrams for the antiferromagnetic long-range Ising model under a linear coupling to a single bosonic mode on the square and triangular lattice.
In the limit of zero coupling, the ground-state magnetization forms a devil's staircase structure of magnetization plateaux as a function of an applied longitudinal field in Ising direction.
Apart from a paramagnetic superradiant phase with a finite photon density at strong linear coupling to a single bosonic mode, the long-range interactions lead to a plethora of intermediate phases that break the translational symmetry of the lattice and have a finite photon density at the same time.
To qualitatively study the ground-state phase diagrams, we apply an adaption of the unit-cell-based mean-field calculations, which capture all possible magnetic unit cells up to a chosen extent.
Further, we exploit an exact mapping of the non-superradiant phases to an effective pure Dicke model in order to calculate upper bounds for phase transitions towards superradiant phases.
Finally, to treat quantum fluctuations in a quantitative fashion, we employ a generalized wormhole quantum Monte Carlo algorithm.
We discuss how these three methods are used in a cooperative fashion.
In the calculated phase diagrams we see several features arising from the long-range interactions: The devil's staircases of distinct magnetically ordered normal phases and non-trivial magnetically ordered superradiant phases beyond the findings for nearest-neighbor interactions.
Examples are a superradiant phase with a three-sublattice magnetic order on the square lattice and the superradiant Wigner crystal with four sites per unit cell on the triangular lattice.
Further, we classify the nature of the quantum phase transitions between different phases.
Concretely, between normal and superradiant phases with the same (different) magnetic order, the transition is of second order with Dicke universality (first order).
Further, between superradiant phases we find first-order phase transitions, besides specially highlighted regimes for which we find indications for continuous second-order behavior.
\end{abstract}

\maketitle

\section{Introduction}

The Dicke model is a paradigmatic model describing the linear coupling of two-level matter degrees of freedom to a single bosonic mode \cite{Dicke1954}.
In the Dicke model, the matter degrees of freedom interact solely with the bosonic mode without any direct interaction between each other \cite{Dicke1954}.
Already in this simple framework, a second-order quantum phase transition from a normal phase with no ground-state photon density to a photon condensate occurs \cite{Hepp1973A,Hepp1973B,Wang1973}.
On the one hand, a natural extension of the Dicke model is to investigate how interacting matter degrees-of-freedom influence the properties of light-matter systems \cite{Zhang2014, Rohn2020, Schellenberger2024, Langheld2024}.
On the other hand, it is interesting how non-trivial ground-states of interacting matter degrees-of-freedom behave in the presence of a coupling to bosonic modes \cite{Mazza2019,Passetti2023,Roche2024A,Weber2023}.

So far, one step towards this research goal was the investigation of the Dicke-Ising model \cite{Zhang2013,Zhang2014,Rohn2020,Schellenberger2024,Schneider2024,Langheld2024}.
Here, the matter degrees-of-freedom are placed on a lattice and are coupled with a nearest-neighbor Ising interaction.
Several analytical and numerical tools to study ground-state phase diagrams of this model have been introduced.
This includes: Classical spin mean-field calculations \cite{Zhang2014}, the mapping of low-energy properties of non-superradiant phases to the pure Dicke-model \cite{Schellenberger2024}, exact diagonalization \cite{Rohn2020,Schuler2020}, and, most notably, fully quantitative quantum Monte Carlo algorithms \cite{Langheld2024}.
The nearest-neighbor antiferromagnetic Ising interaction leads to the existence of structured non-superradiant phases \cite{Zhang2014,Langheld2024}, as well as, intermediate phases with a non-trivial ordering momentum in the matter entities and a finite boson density \cite{Zhang2014,Langheld2024}.
Of course, there are numerous possible extensions of the paradigmatic nearest-neighbor Dicke-Ising model, e.\,g., changing the interaction between the matter entities, introducing quenched disorder in the couplings, or altering the mode structure of the bosonic degree(s)-of-freedom.

Another natural extension are algebraically decaying long-range Ising interactions between the matter degrees-of-freedom on the lattice.
In general, a lattice model with repulsive algebraically decaying Ising interactions with a decay exponent $\alpha>d$ with $d$ the dimension of the system and a longitudinal field leads to a ground-state phase diagram containing devil's staircase structures of magnetization plateaux \cite{Koziol2023,Koziol2024}. 
One theoretically interesting aspect in this setting is to investigate how the infinite number of non-superradiant crystalline phases in the devil's staircase of the matter Hamiltonian melt under the coupling to the single bosonic mode.

In general, such algebraically decaying long-range interactions occur in numerous state-of-the-art quantum simulation platforms, where the relevant degrees of freedom can be arranged in a lattice \cite{Bloch2005,Bloch2008,Houck2012,Zhang2014,Barredo2016,Barredo2018,OhldeMello2019,Browaeys2020,Su2023}.
On the one hand, interactions decaying with $r^{-3}$ are realized using static magnetic atomic dipoles like in ultracold trapped Erbium atoms in optical lattices \cite{Su2023} or between Rydberg atoms prepared in dipole-coupled Rydberg states \cite{Browaeys2020}.
On the other hand, van-der-Waals interactions with $r^{-6}$ are realized in Rydberg atom systems, where the same Rydberg level is excited in two atoms and the dipole interaction is a second-order process \cite{Browaeys2020}.
Although an untruncated algebraically decaying long-range interaction seems unfeasible in the current state of superconducting qubit quantum simulators, truncated long-range interactions are well within reach depending on the existing hardware limitations \cite{Wallraff2004,Johnson2011,Viehmann2011,Zhang2014,Boothby2015,Rosenberg2017,Zhao2021,Derakhshandeh2024}.
To achieve progress in this direction, the greatest challenge is to increase the connectivity of the qubits, which aligns with one of the main research goals for superconducting circuit quantum computing \cite{Rosenberg2017,Zhao2021,Riel2021,Derakhshandeh2024}.
A promising way to overcome connectivity issues in this type of systems is a three-dimensional integration of the circuits \cite{Rosenberg2017,Zhao2021,Derakhshandeh2024}.

These experimental platforms result in a second motivation to study the Dicke model with long-range Ising interactions, since they can be easily placed in cavity/circuit quantum electrodynamic setups.
Realizing superradiant phases with a finite ground-state photon density in equilibrium systems is a highly debated topic in literature with fundamental no-go theorems for many proposed setups \cite{Rzazewski1975,Bialynicki1979,Gawedzki1981,Rzazewski1991,Keeling2007,Nataf2010A,Nataf2010B,Vukics2012,Viehmann2011,Bamba2016,Andolina2019,Andolina2022}.
Therefore, regarding the current state of the literature, the common way to realize Dicke-type models is as an effective description of non-equilibrium systems \cite{Dimer2007,Baumann2010,Bastidas2012,Gelhausen2016,Zhiqiang2017,Zhang2018,Puel2024}.
For the model described theoretically in this work, we expect a Rydberg-dressed spin lattice coupled to a single cavity mode \cite{Gelhausen2016,Puel2024} via cavity-assisted Raman transitions \cite{Dimer2007,Zhiqiang2017,Zhang2018} to be the most promising platform to realize the Dicke model with long-range Ising interactions.

A third argument that motivates the study of the long-range Dicke Ising model are the current developments in the numerical treatments of systems with long-range interactions \cite{Sandvik2003,Sun2017,Fey2019,Koziol2023,Adelhardt2024,Koziol2024}, as well as recent (semi)-analytical insights to the low-energy physics of non-superradiant phases in extended Dicke models \cite{Schellenberger2024}.
In particular, we will use a unit-cell based optimization approach to extend the mean-field calculations for the nearest-neighbor Dicke-Ising model to full, untruncated long-range interactions \cite{Koziol2023,Koziol2024}.
Further, we will demonstrate how to use the resummed couplings from the unit-cell based optimization approach to extend the mapping of the low-energy properties of non-superradiant states to the Dicke model from the nearest-neighbor Dicke-Ising model to models with long-range interactions \cite{Schellenberger2024}.

For the sake of this work, we focus on the long-range Dicke Ising model on the square and triangular lattice.
We choose to study two-dimensional lattices motivated by atomic experimental platforms that often investigate two-dimensional geometries for convenient readout with quantum gas microscopes and cameras.
Another reason is that two-dimensional geometries are easier to fabricate on a chip in the context of a realization in coupled superconducting qubits.
For demonstrations we choose dipolar and van-der-Waals Ising interactions.
Generalizations of all the methods used in this work to arbitrary other lattices of interest are straightforward.

This paper is structured as follows: 
In Sec.~\ref{sec:LRDIM} we formally introduce the long-range Dicke-Ising model and provide a comprehensive review of the limiting cases already studied in the literature.
Then, in Sec.~\ref{sec:UCMF} we present how mean-field calculations can be implemented for the long-range Dicke-Ising model using the unit-cell based optimization scheme introduced in Refs.~\cite{Koziol2023,Koziol2024}.
Further, in Sec.~\ref{sec:AlmostEverything} we discuss how to extend the mapping to the pure Dicke model \cite{Schellenberger2024} for arbitrary magnetic orders contained in the devil's staircase.
We introduce the generalized wormhole quantum Monte Carlo scheme \cite{Langheld2024} for the long-range Dicke Ising model in Sec.~\ref{sec:MonteCarlo}, before finishing the section on methods with an outline how we use all three complementary approaches in a joint fashion in Sec.~\ref{sec:ComplementaryApproaches}.
The results for the square lattice are presented in Sec.~\ref{sec:ResultsSquare} while the results for the triangular lattice are presented in Sec.~\ref{sec:ResultsTriangular}.
We conclude and summarize our analysis in Sec.~\ref{sec:summaryandoutlook} and provide a brief outlook.

\section{Long-range Dicke Ising model}
\label{sec:LRDIM}
In this work, we consider ground-state properties of the long-range Dicke Ising model (LRDIM).
This model combines the Dicke model of $N$ spins-$1/2$ uniformly coupled with strength $g/\sqrt{N}$ to a single bosonic mode with frequency $\omega>0$ and an algebraically decaying long-range Ising interaction between the spins.
The Hamiltonian reads
\begin{align}
\begin{split}
		\label{eq:LRDIM}
		H =&\ \frac{\epsilon}{2}\sum_i\sigma_i^z+\frac{1}{2}\sum_{i\neq j}\underbrace{\frac{J}{|\vec r_i - \vec r_j|^\alpha}}_{J_{i,j}}\sigma_i^z\sigma_j^z \\
		   &\ +\omega a^\dagger a + \frac{g}{\sqrt{N}}\left(a+a^{\dagger}\right)\sum_i \sigma_i^x
\end{split}
\end{align}
with Pauli matrices $\sigma_i^{x/z}$ describing spins at positions $\vec r_i$ and $a^{(\dagger)}$ describing annihilation (creation) operators of the bosonic mode.
The longitudinal field is parametrized by $\epsilon$ and the long-range Ising interaction has the amplitude $J$ and decays with the distance between the spins to the power of $\alpha$.
Note, the nearest-neighbor limit is recovered by $\alpha\rightarrow\infty$.
For the sake of this work, we consider antiferromagnetic Ising interactions ($J>0$) and weak long-range interactions \cite{Defenu2023} with $\alpha>d$ where $d$ is the spatial dimension of the system.
Further, we focus on the two-dimensional square and triangular lattice.
The square lattice is a prime example for a bipartite lattice, while the triangular lattice is non-bipartite and can host frustration effects in the limit of antiferromagnetic nearest-neighbor interactions.
Both lattices have a primitive unit cell with one site.

In the nearest-neighbor limit, the ground-state properties of the Dicke Ising model (DIM) have been studied in numerous publications on bipartite lattices \cite{Zhang2013,Zhang2014,Rohn2020,Puel2024,Schellenberger2024,Schneider2024,Langheld2024}.
In general, ground states are characterised to be normal (N) if the photon density $n_{\text{ph}}=\langle a^\dagger a\rangle/N$ in the ground state is zero and superradiant (SR) if it is finite.
For the linear chain, the established picture is that for $\epsilon=0$ there is a first-order phase transition between a magnetically ordered N phase at small $g$ values and a paramagnetic SR phase at large $g$ values.

For ferromagnetic Ising interactions the two phases remain for $\epsilon>0$ and the point of the phase transition $g_c$ moves to larger $g$ values \cite{Langheld2024}. 
For small $\epsilon$ values, the phase transition remains of first order, but for larger $\epsilon$ values the transition is in the second order Dicke universality \cite{Langheld2024}.

Regarding antiferromagnetic Ising interactions the system has been studied on bipartite lattices by a mean-field ansatz, treating the spins as classical vectors \cite{Zhang2014}, and quantitive wormhole quantum Monte Carlo (QMC) simulations \cite{Langheld2024}.
The qualitative mean-field results from Ref.~\cite{Zhang2014} suggest an antiferromagnetic N phase at small $\epsilon$ and $g$ values, that is enclosed by an antiferromagnetic SR phase, which is on the other hand enclosed by a paramagnetic SR phase at large values of $g$.
Further, at $g=0$, there is a first-order phase transition to a paramagnetic N phase for a given $\epsilon$ value \cite{Zhang2014}.
The paramagnetic N phase breaks down with a second-order phase transition to the paramagnetic SR phase \cite{Zhang2014}.
For increasing $\epsilon$ the critical point shifts to larger values of $g$ and converges to the expression for the Dicke model $g_c = \sqrt{\omega \epsilon}/2$.
In the mean-field picture \cite{Zhang2014} only the phase transitions on the $\epsilon=0$ and $g=0$ axis are of first order, all other phase transition lines are of second order.
The quantitative QMC simulation analysis \cite{Langheld2024} confirms the second-order mean-field phase transition line between the paramagnetic N phase and the paramagnetic SR phase.
Contrary to the mean-field analysis, the QMC shows that the first-order phase transition between the antiferromagnetic N phase and the paramagnetic SR phase for $\epsilon=0$ actually extends to finite $\epsilon$.
Therefore, the antiferromagnetic N phase is no longer fully
\footnote{To be precise, even in the mean-field calculations there is no antiferromagnetic SR phase in the phase diagram for the $\epsilon$-axis at $g=0$, as well as, the $g$-axis for $\epsilon=0$ \cite{Zhang2013,Zhang2014,Rohn2020,Langheld2022}.} 
enclosed by the antiferromagnetic SR phase. 
The intermediate antiferromagnetic SR phase is proven to be present in one and two dimensions at moderate values of $\epsilon$, with a second-order Dicke criticality transition to the antiferromagnetic N phase.
According to Ref.~\cite{Langheld2024}, on the square lattice the non-trivial transition line between the antiferromagnetic and paramagnetic superradiant phase (AS and PS), which emerges from the first-order transition line between antiferromagnetic N and paramagnetic SR phase, remains of first order close to the split-up, but changes to a continuous quantum phase transition with 3D Ising criticality for smaller values of $g$.
On the linear chain, this transition was found to remain of first-order in the investigated parameter region \cite{Langheld2024}.

Besides the nearest-neighbor limit, the ground states of the pure spin model for $g=0$ have been studied in the literature \cite{Koziol2023,Koziol2024}.
In this limit, the model reduces to a classical antiferromagnetic long-range Ising model in a longitudinal magnetic field \cite{Koziol2023,Koziol2024}.
The ground states of this model for different $\epsilon$ values are given by a devil's staircase structure of magnetization plateaux \cite{Koziol2023,Koziol2024}.
For $\epsilon=0$, the system has a lattice-dependent ground state with zero magnetization.
For the square lattice, this is the antiferromagnetic state known from the nearest-neighbor limit \cite{Koziol2024} and for the triangular lattice those are plain stripes \cite{Koziol2019,Koziol2023,Koziol2024}.
For increasing $\epsilon$ one goes through the plateaux of the devil's staircase until the fully field-polarized state is reached.

Further, it has been shown for models like the LRDIM \eqref{eq:LRDIM} that the $g=0$ product-state ground states remain the exact ground states until a phase transition occurs \cite{Rohn2020,Schellenberger2024}.
This includes the case for all magnetization plateaux of the devil's staircase.

\section{Methods}
\label{sec:methods}
In order to study the LRDIM and derive a quantitative ground-state phase diagram, we employ three distinct complementary approaches.
First, we apply a unit-cell-based mean-field optimization approach (see Sec.~\ref{sec:UCMF}) which provides numerically exact results in the limit of $g=0$ and draws a qualitative phase diagram in the entire parameter space.
Second, the mapping of normal phases to the Dicke model for systems with long-range Ising interactions is discussed in Sec.~\ref{sec:AlmostEverything}. 
With this method the ground-state properties and elementary excitations can be studied out of the $g=0$ limit in an exact fashion for the thermodynamic limit.
If the true transition point coincides with the gap closing of the effective Dicke model, the transition is of Dicke criticality. Otherwise the condensation provides an upper bound for a phase transition in the light-matter coupling $g$. 
Third, we describe a generalized wormhole quantum Monte Carlo to evaluate the quantum phase diagram quantitatively (see Sec.~\ref{sec:MonteCarlo}).
We conclude this section with a discussion on how these three approaches complement each other and how the results of each method aid the other approaches.

\subsection{Unit-cell-based mean-field optimization}
\label{sec:UCMF}
In this work, we extend the approach developed in Refs.~\cite{Koziol2023,Koziol2024} to perform mean-field calculations \cite{Zhang2014} for the LRDIM. 
The main idea is to treat the spins as classical vectors of length one parametrized by angles $\phi_i$ and to describe the bosonic mode as a coherent state with parameter $\lambda_0 \in \mathds{R}$.
The parametrization of the spins in $xz$-plane reads
\begin{align}
	\label{eq:ClassicalSpinApprox}
	\sigma_i^x \rightarrow \cos(\phi_i) \hspace{0.5cm} \text{and} \hspace{0.5cm} \sigma_i^z \rightarrow \sin(\phi_i) \ .
\end{align}
For the coherent state we use
\begin{align}
	\label{eq:CoherentState}
	\ket{\lambda_0}=\exp(-\frac{\lambda_0^2}{2}+\lambda_0 a^{\dagger}) \ket{0} \ ,
\end{align}
which is a right eigenstate to the annihilation operator.
Performing the classical spin approximation~\eqref{eq:ClassicalSpinApprox} for the LRDIM~\eqref{eq:LRDIM} and taking the expectation value w.\,r.\,t. the coherent state Eq.~\eqref{eq:CoherentState} gives the following classical energy function
\begin{align}
\begin{split}
		E(\{J_{i,j}\}, \epsilon, g, \omega;\, & \lambda_0, \{\phi_{i}\}) = \phantom{\bigotimes_{i=j}} \\
		&\ \frac{\epsilon}{2}\sum_i\sin(\phi_i) +\sum_{i\neq j} \frac{J_{i,j}}{2}\sin(\phi_i)\sin(\phi_j)\\
																	 & \ +\omega \lambda_0^2 + 2\frac{g}{\sqrt{N}}\lambda_0 \sum_i\cos(\phi_i)
\end{split}
\end{align}
dependent solely on the parameters of the Hamiltonian and the variables $\lambda_0$ and $\{\phi_{i}\}$ defining the state.

To determine the ground state of the classical energy function we apply the approach developed in \mbox{Refs.~\cite{Koziol2023,Koziol2024}}.
We consider systematically all possible unit cells up to a certain extent and treat the long-range interaction on each unit cell using appropriately resummed interactions (see Eqs.~\eqref{eq:resummendbegin}~to~\eqref{eq:resummendend}) \cite{Koziol2023, Koziol2024}, absorbing the interactions between translations of the unit cell into the couplings within the cell.
With these resummed couplings the energy of classical periodical Ising configurations can be evaluated in the thermodynamic limit solely on their unit cell \cite{Koziol2023}.
For $g=0$, this treatment of the long-range interactions provides the exact energies of states in the thermodynamic limit by considering only the unit cell of a state \cite{Koziol2023,Koziol2024}.
It then transfers to an exact treatment of the long-range Ising interactions on a mean-field level \cite{Koziol2024}.
Having the mean-field Hamiltonian on each unit cell with the resummed couplings, we determine the optimal configuration on each unit cell for a given set of parameters \cite{Koziol2023, Koziol2024}.
In the end, we compare the energies per site between each unit cell to determine the overall optimal configuration minimizing the energy \cite{Koziol2023, Koziol2024}.

The key insight making this approach possible is that one can rewrite a diagonal -- in our case Ising -- long-range interaction for a periodic pattern with a $K$-site unit cell with translational vectors $\vec T_1$ and $\vec T_2$ 
\begin{align}
		\frac{1}{2}\sum_{i\neq j}\frac{J}{|\vec r_i -\vec r_j|^\alpha}\sigma_i^z\sigma_j^z=\frac{1}{2}\sum_{i=1}^K \bar J^{\alpha}_{i,i}\sigma_i^z\sigma_i^z + \frac{1}{2}\sum_{i\neq j}^K \bar J^{\alpha}_{i,j}\sigma_i^z\sigma_j^z
\end{align}
as sums over the unit cell of the configuration. 
The appropriately resummed couplings are
\begin{align}
		\label{eq:resummendbegin}
		\bar J^{\alpha}_{i,j} & = J \sum_{k=-\infty}^{\infty}\sum_{l=-\infty}^{\infty} \frac{1}{|\vec r_i -\vec r_j + l\vec T_1 - k \vec T_2|^\alpha}\\
						 & = J \sum_{\vec u \in \Lambda(\vec T_1, \vec T_2)} \frac{1}{|\vec r_i -\vec r_j + \vec u|^\alpha} \\
						 & = J \ \zeta_{\Lambda(\vec T_1, \vec T_2), \alpha}(\vec r_i - \vec r_j, \vec 0) \\
						 \nonumber & \\
						 \bar J^{\alpha}_{i,i} & = J \sum_{k=-\infty}^{\infty}\sum_{l=-\infty}^{\infty} \frac{(1-\delta_{l,0}\delta_{k,0})}{| l\vec T_1 - k \vec T_2|^\alpha} \\
						 & = J \sum_{\vec u \in \Lambda(\vec T_1, \vec T_2){\setminus \vec 0}} \frac{1}{|\vec u|^\alpha} \\
						 \label{eq:resummendend}
						 & = J \ \zeta_{\Lambda(\vec T_1, \vec T_2), \alpha}(\vec 0, \vec 0)
\end{align} 
with $\delta_{i,j}$ being the Kronecker delta, $\Lambda(\vec T_1, \vec T_2)$ the lattice spanned by $\vec T_1$ and $\vec T_2$, and $\zeta_{\Lambda, \alpha}(\vec x, \vec y)$ being the Epstein $\zeta$-function \cite{Epstein1903,Epstein1906}.
Note, for the sums to converge, we require $\alpha$ to be larger than the dimension of the lattice $d$.
The Epstein $\zeta$-function can be efficiently calculated for lattices in arbitrary dimensions \cite{Crandall2012,Buchheit2021, Buchheit2022, Buchheit2024, Buchheit2024Code, Buchheit2024Code1}. Therefore the considered resummed couplings can be considered exact up to machine precision.

The next step is to formulate an expression for the energy per site for each unit cell.
For a $K$-site unit cell we obtain for the energy density $e=E/N$
\begin{align}
\begin{split}
		e(\{\bar J_{i,j}\},& \epsilon, g, \omega;\,  \lambda_0, \{\phi_{i}\}) = \phantom{\bigotimes_{i=j}} \\
		&\ \frac{\epsilon}{2K} \sum_i\sin(\phi_i) + \frac{1}{K} \sum_{i, j} \frac{\bar J^{\alpha}_{i,j}}{2}\sin(\phi_i)\sin(\phi_j)\\
																	 & \ +\omega \lambda + \frac{2g\lambda}{K} \sum_i\cos(\phi_i)
\end{split}
\end{align}
with $\lambda=\lambda_0/\sqrt{N}$.

On each unit cell, we search for the energetically optimal arrangement of angles $\phi_i$ and the coherent state parameter $\lambda$ with a global optimization scheme \cite{Jones1993,Gablonsky2001}, as well as local low-storage Broyden-Fletcher-Goldfarb-Shanno algorithm \cite{Broyden1970,Fletcher1970,Goldfarb1970,Shanno1970} for numerous relevant starting configurations.
In the end, we compare the energy per site of the optimal states on each unit cell, to determine the mean-field ground state for the thermodynamic limit.

The last remaining step in the procedure is to characterize the ground state from the optimal angles $\phi_i$ and the coherent state parameter $\lambda$. 
If $\lambda > 0$ we call a state superradiant (SR) and if $\lambda = 0$ the state is normal (N).
Further, we label the spin states by the fractions of sites having the same $\sin(\phi_i)$ value. 
For a state with $l$ different values we therefore list $l$ fractions, starting from the smallest value of $\sin(\phi_i)$, and going to the largest one.
Following our nomenclature, the phases described in \cite{Zhang2014,Langheld2024} become the following: antiferromagnetic N phase $\rightarrow$ $\frac{1}{2}$-$\frac{1}{2}$ N phase; paramagnetic N phase $\rightarrow$ $\frac{1}{1}$ N phase; antiferromagnetic SR phase $\rightarrow$ $\frac{1}{2}$-$\frac{1}{2}$ SR phase; paramagnetic SR phase $\rightarrow$ $\frac{1}{1}$ SR phase.

Using the angles $\phi_i$ and the coherent state parameter $\lambda$ also observables like magnetizations and static structure factors can be calculated using Eq.~\ref{eq:ClassicalSpinApprox}, as well as, e.\,g., the mean photon density.

For the analysis of the square lattice presented in this work we use the following definitions:
We choose the primitive translation vectors to be $\vec t_1^{\ \square}=(1,0)^\text{T}$ and $\vec t_2^{\ \square}=(0,1)^\text{T}$. 
For the optimization we use all distinct unit cells with translation vectors $\vec T_1, \vec T_2\in\mathcal{A}_6^{\square}$ out of the set as in \cite{Koziol2023,Koziol2024}
\begin{align}
\label{eq:setsquare}
\mathcal{A}_6^{\square} = \{i\vec t_1^{\ \square}+j\vec t_2^{\ \square}|i\in\{-6,...,6\},j\in\{-6,...,6\}\} \ .
\end{align}

For the analysis of the triangular lattice presented in this work we use the following definitions:
We choose the primitive translation vectors to be $\vec t_1^{\ \triangle}=(1,0)^\text{T}$ and $\vec t_2^{\ \triangle}=(1/2,\sqrt{3}/2)^\text{T}$. 
For the optimization we use all distinct unit cells with translation vectors $\vec T_1, \vec T_2\in\mathcal{B}_6^{\triangle}$ out of the set as in \cite{Koziol2023,Koziol2024}
\begin{align}
\label{eq:settriangular}
\mathcal{B}_6^{\triangle} = &\{i\vec t_1^{\ \triangle}+j\vec t_2^{\ \triangle}|\\
		\nonumber & \qquad i\in\{-6,...,6\},\\ 
\nonumber & \qquad j\in\{\max(-6-i,-6),...,\min(6-i,6)\}\\
		  \nonumber&\} \ .
\end{align}

\subsection{Mapping non-superradiant phases to the Dicke model}
\label{sec:AlmostEverything}

In addition to the mean-field analysis described above, we also follow the findings of Ref.~\cite{Schellenberger2024} showing that it is possible to map non-SR correlated light-matter systems to the exactly solvable Dicke model.
In this work, we apply the procedure of Ref.~\cite{Schellenberger2024} to the devil's staircase of N phases at $g=0$ and determine the effective Dicke model within the N phases.
This effective Dicke model is solvable by a bosonic Bogoliubov transformation in the $\vec k=\vec 0$ subspace \cite{Hepp1973A,Emary2003}.
By investigating the breakdown of the Bogoliubov transformation, we can determine potential second-order quantum phase transitions between N and SR phases \cite{Schellenberger2024}.

In the following, we describe the calculation for an arbitrary normal ground state with a $K$-site unit cell.
That means, this procedure will be repeated for each $\epsilon$ value in the $g=0$ devil's staircase.
The first step is to perform sublattice rotations $\sigma_i^z\rightarrow-\sigma_i^z$ in the state and Hamiltonian~\eqref{eq:LRDIM} in such a way that the ground state is fully polarized with all spins point in negative $z$-direction with $\sigma_i^z=-1$. 
In the Hamiltonian we denote this by phases $\theta_i\in\{-1,1\}$ for each lattice site dependent on the $g=0$ state
\begin{align}
\begin{split}
		\label{eq:LRDIMRotated}
		H =&\ \frac{\epsilon}{2}\sum_i\theta_i\sigma_i^z+\frac{1}{2}\sum_{i\neq j}J_{i,j}\theta_i \theta_j\sigma_i^z\sigma_j^z \\
		   &\ +\omega a^\dagger a + \frac{g}{\sqrt{N}}\left(a+a^{\dagger}\right)\sum_i \sigma_i^x \ .
\end{split}
\end{align}
Next, we bring Eq.~\eqref{eq:LRDIMRotated} into a hardcore-bosonic picture using the Matsubara-Matsuda transformation \cite{Matsubara1956}
\begin{align}
		\sigma_i^z = 2b_i^\dagger b_i^{\phantom{\dagger}}-1 \hspace{1cm} \sigma_i^x=b_i^\dagger+b_i^{\phantom{\dagger}}
\end{align}
with $b_i^{\phantom{\dagger}}$ ($b_i^\dagger$) being hardcore bosonic annihilation (creation) operators.
Inserting the transformation, the resulting Hamiltonian reads 
\begin{align}
\begin{split}
		\label{eq:LRDIMRotatedBosons}
		H =&\ -\frac{\epsilon}{2} \sum_i\theta_i + \frac{1}{2}\sum_{i\neq j}J_{i,j}\theta_i \theta_j \\
		   &\ +\omega a^\dagger a + \frac{g}{\sqrt{N}}\sum_i\left(a+a^{\dagger}\right) \left(b_i^\dagger+b_i^{\phantom{\dagger}}\right) \\
		   &\ +\sum_i\theta_i \, \left(\epsilon-2\sum_{j\neq i} J_{i,j}\theta_j\right)\, b_i^\dagger b_i^{\phantom{\dagger}} \\
		   &\ +2\sum_{i\neq j}J_{i,j}\, \theta_i \theta_j \, b_i^\dagger b_i^{\phantom{\dagger}} b_j^\dagger b_j^{\phantom{\dagger}} \ .
\end{split}
\end{align}
The key insight of the mapping described in Ref.~\cite{Schellenberger2024} is that the quartic term in the hardcore bosonic operators in Eq.~\eqref{eq:LRDIMRotatedBosons} decouples in the thermodynamic limit from the light-matter interaction for $\sum_{i \neq j}|J_{i,j}\, \theta_i \theta_j|<\infty$ which certainly holds for $\alpha>d$ in the LRDIM.
The essence of the underlying argument is that the light couples only to the zero-momentum mode, whose overlap with the matter-matter interactions scales with $1/N$ and therefore vanishes in the thermodynamic limit. 

Let us introduce the Fourier transformed magnon operators
\begin{align}
		\tilde b_{\vec k,\beta}^{\phantom{\dagger}}&=\sqrt{\frac{K}{N}}\sum_{\vec u \in\Lambda(\vec T_1, \vec T_2)} b_{\vec u, \beta} e^{-{\rm i}\vec k \vec u} \\
		\tilde b_{\vec k,\beta}^{\dagger}&=\sqrt{\frac{K}{N}}\sum_{\vec u \in\Lambda(\vec T_1, \vec T_2)} b^{\dagger}_{\vec u, \beta} e^{{\rm i}\vec k \vec u} 
\end{align}
where $\beta$ labels the sites within the unit cell.
The light-matter interaction transforms then to
\begin{align}
		\frac{g}{\sqrt{K}}\sum_{\beta}\left(a+a^{\dagger}\right)\left(\tilde b_{\vec 0, \beta}^{\phantom{\dagger}}+\tilde b_{\vec 0, \beta}^{\dagger}\right) 
\end{align}
and the resulting relevant low-energy Dicke model \cite{Schellenberger2024} reads
\begin{align}
\begin{split}
		\label{eq:LRDIMDicke}
		H =& \ +\omega a^\dagger a \\
		   & \ + \frac{g}{\sqrt{K}}\sum_{\beta}\left(a+a^{\dagger}\right) \left(\tilde b_{\vec 0, \beta}^\dagger+\tilde b_{\vec 0, \beta}^{\phantom{\dagger}}\right) \\
		   & \ + \sum_{\beta}\theta_\beta (\epsilon-2\sum_{\gamma} \bar J_{\beta,\gamma}^{\alpha}\theta_{\gamma})  b_{\vec 0, \beta}^\dagger b_{\vec 0, \beta}^{\phantom{\dagger}} \ .
\end{split}
\end{align}
With the Dicke model \eqref{eq:LRDIMDicke} we study the low-energy physics in the normal phases.
This is done by diagonalizing the Hamiltonian using a bosonic Bogoliubov transformation \cite{Xiao2009,Bogoliubov1947,Bogoliubov1958A,Bogoliubov1958B,Valatin1958}, where Eq.~\eqref{eq:LRDIMDicke} is rewritten as a symmetric matrix and brought in a dynamical matrix form \cite{Xiao2009,Schellenberger2024} to ensure the particle-statistics-preserving nature of the transformation.
For the technical details we refer to Ref.~\cite{Schellenberger2024}.
From the diagonalisation of the dynamical matrix we can, for example, determine the position of second-order phase transitions.
To do this we investigate the spectrum of the dynamical matrix and if at least one eigenvalue has a vanishing real part, the normal phase is no longer the ground state of the system.
Sweeping from $g=0$ to the breakdown point $\bar g$ therefore gives an upper bound for the extent of the normal phase.
Of course this treatment of the normal phases is not capturing first-order phase transitions between states of different symmetry.
If a transition in the full LRDIM without any approximation coincides with the phase transition point of this method, it is clear that the phase transition is a second-order phase transition in the Dicke universality class \cite{Schellenberger2024,Langheld2024}.

\subsection{Generalized wormhole quantum Monte Carlo}
\label{sec:MonteCarlo}
In addition to the mean-field analysis and mapping of the non-SR light-matter phases, we utilize a recently introduced wormhole quantum Monte Carlo method for Dicke-spin systems \cite{Langheld2024}, essentially following the implementation instructions in \cite{Langheld2024}. In the following we will explain the basic idea of this method and, in particular, the adaptions made for this paper to accomodate for long-range Ising interactions.
This QMC method is able to quantitatively determine ground-state properties of mesosopic spin systems coupling to common bosonic modes.
In comparison to the mapping of the non-SR light-matter phases, this method can therefore also determine first-order phase transitions due to a level crossing but is limited by the computational effort one wants to invest.

The method is based on the wormhole algorithm \cite{Weber2017,Weber2022}, in which bosonic degrees of freedom are integrated out analytically prior to the simulation.
For spin-boson systems like the Dicke-Ising model this leads to an effective spin model with imaginary-time retarded interactions. 
In comparison to the original wormhole algorithm \cite{Weber2017, Weber2022}, the algorithm specifically introduced for Dicke-spin systems \cite{Langheld2024} addresses the dissimilarity between the intrinsic spin-spin and the photon-induced, mean-field like spin-spin interactions.

The partition function of the effective spin model is given by
\begin{equation}
	\label{eq:ret-part-func}
	Z = Z_{\mathrm{b}} \Tr_{\mathrm{s}}\left[ \mathcal{T}_{\tau} e^{-{S}_{\mathrm{ret}}} \right]
\end{equation}
with $Z_{\mathrm{b}}$ the partition function of the free bosons, $\mathcal{T}_{\tau}$ the time-ordering operator for imaginary time and $\Tr_{\mathrm{s}} \left[\dots\right]$ the trace over the Hilbert space of the spins.
The retarded spin-spin action is of the form
\begin{align}
	S_{\mathrm{ret}} =& \int_0^\beta \dd{\tau} H^{\mathrm{spin}}(\tau) + \int_0^\beta \int_0^\beta \dd{\tau}\dd{\tau'} H^{\mathrm{ret}}(\tau, \tau')
\end{align}
with $H^\mathrm{spin}$ the pure spin part of the Hamiltonian and $H^{\mathrm{ret}}$ the imaginary-time retarded part mediated by the bosonic mode.
For the long-range Dicke-Ising model, they are given by
\begin{align}
	H^{\mathrm{spin}}(\tau) &= \frac{\epsilon}{2} \sum_i \sigma^z_i(\tau) + \sum_{i,j} \frac{J_{i,j}}{2} \sigma^z_i(\tau) \sigma^z_j(\tau)\\
	H^{\mathrm{ret}}(\tau, \tau') &= -\sum_{i,j} \frac{g^2}{\omega N} \sigma^x_i(\tau) P_{\mathrm{sym}}(\omega, \abs{\tau - \tau'}) \sigma^x_j (\tau') 
\end{align}
where $P_{\mathrm{sym}}(\omega, \Delta\tau) = \frac{\omega}{2} \frac{e^{-(\beta - \Delta\tau) \omega} + e^{-\Delta\tau \omega}}{1- e^{-\beta\omega}}$ is the (normalized) symmetric free boson propagator.

The partition function of the effective spin model is then expanded as in the SSE approach \cite{Sandvik1991, Sandvik2010}
\begin{align}
	\frac{Z}{Z_b} &= \sum_{\{\ket{\alpha}\}} \sum_{n=0}^{\infty} \sum_{\{C_n\}}\frac{1}{n!} \bra{\alpha} \mathcal{T}_\tau  \prod_{k=1}^n H_{\nu_k} \dd{\tau_k}\dd{\tau'_k}\ket{\alpha} \label{eq:part-func-expansion}
\end{align}
with a suitable computational basis $\{\ket{\alpha}\}$ and decomposition $S_{\mathrm{ret}} = -\SumInt_{\nu} H_\nu \dd{\tau}\dd{\tau'}$ into non-branching bond operators $H_\nu$ \cite{Sandvik1991, Sandvik1997} with vertex variables \mbox{$\nu = \{\mathrm{type}, i,j, \tau, \tau'\}$}.

In the case of the Dicke-Ising model, the computational basis is chosen to be the $\sigma^z$-eigenbasis such that the spin part of the Hamtilonian is diagonal and the sign problem can be avoided even for antiferromagnetic non-bipartite lattices.
The action is decomposed into
\begin{align}
	H^{\mathrm{spin}}_{i,j}(\tau, \tau') =& \frac{\delta_{\tau, \tau'}}{2} \left[C_{i,j} - J_{i,j} \sigma^z_i(\tau) \sigma^z_j (\tau) \right.\\ 
		&- \frac{\abs{J_{i,j}}}{\sum_j \abs{J_{i,j}}} \frac{\epsilon}{2} \left. (\sigma^z_i(\tau) + \sigma^z_j(\tau))\right]\\
	H^{\mathrm{ret}}_{i,j}(\tau, \tau') =& \frac{g^2}{\omega N} \sigma^x_i(\tau) P_{\mathrm{sym}}(\omega, \abs{\tau - \tau'}) \sigma^x_j (\tau') 
\end{align}
with the artificially introduced constant $C_{i,j}=\abs{J_{i,j}} +  \abs{\epsilon}\abs{J_{i,j}}/\sum_j \abs{J_{i,j}}$ to ensure non-negativity of all matrix elements of the diagonal operators $H^{\mathrm{spin}}_{i,j}(\tau, \tau')$ in the computational basis.

While in the algorithm for the nearest-neighbor Dicke-Ising model \cite{Langheld2024} the strength of the longitudinal field $\epsilon$ is distributed equally among the bonds of each spin, in the present long-range case we choose to distribute the longitudinal field strength $\epsilon$ according to the weight $J_{i,j}$ of the Ising bond such that the weight of the diagonal operators connecting two spins $i$ and $j$ is proportional to their Ising coupling strength $\abs{J_{i,j}}$.

With this setup one can now apply the wormhole algorithm with the same solution of the directed-loop equations as in Ref.~\cite{Langheld2024} as the algorithm for the short-range Dicke-Ising model is already agnostic of the lattice geometry and therefore straightforwardly generalizes to the long-range counterpart.
The main difference is that, whenever drawing a diagonal operator connecting two spins on a bond, the probability to draw a certain bond is not the same for all bonds but is proportional to the Ising coupling $~ J_{i,j}$ of the spins $i,j$ connected by the bond.
The first spin $i$ of a bond can be drawn uniformly from all spins due to translational invariance, while the second spin $j$ is then efficiently drawn with relative weight $\sim J_{ij}$ using a Walker alias \cite{Fukui2009}.
Other than the usage of a Walker alias to draw a spin according to the long-range Ising interaction strength, the implementation is analogous to Ref.~\cite{Langheld2024}.

As the expected ordering patterns of the solid N phases are quite diverse and can be arbitrarily complex, it is important to seed the start configuration with the expected spin ordering pattern(s).
Otherwise one will get patterns of solids with many local defects due to several local growths of the pattern which are incompatibly sheered or twisted against each other.
We therefore use a swipe mechanism for the simulations, where we start the simulation for a fixed parameter set $\epsilon, J, \omega,$ with a phase transition at $g_c$ for spin-photon coupling $g_l < g_c$ ($g_h > g_c$) with the expected ordering pattern and increase (decrease) the spin-photon coupling $g$ gradually until $g > g_c$ ($g < g_c)$ going past the phase boundary. 
Sometimes it might speed up the convergence by starting the simulation deep in a phase before performing the actual simulation in the vicinity of the critical point.

For first-order transitions this gives an energy-level crossing of the two respective ground states at $g=g_l$ and $g=g_h$. 
This level crossing is calculated for several system sizes from $12\times 12$ up to $48 \times 48$ with periodic boundary conditions and resummed couplings \cite{Fukui2009,FloresSola2015,Koziol2021,Zhao2023,Song2024,Adelhardt2024} and extrapolated to the thermodynamic limit in $1/N$. 
The specific system sizes used for the simulation depend on the expected phases as the ordering patterns are not compatible with every system size, e.\,g., for a transition between phases with $\frac{3}{5}$-$\frac{2}{5}$ and $\frac{1}{2}$-$\frac{1}{2}$ ordering patterns we take the linear system size $L$ to be a multiple of 10.
For second-order transitions, one phase is adiabatically tuned into the other phase and the transition point can be extracted by a data collapse of the squared order parameter.
In the case of transitions from normal to superradiant phase, we use the photon density $n_{\mathrm{ph}}$, and in the case of magnetic ordering transitions one could use the squared magnetization $m^2(\vec k)$ at the ordering wavevector $\vec k$.

\subsection{Complementary application of the approaches}
\label{sec:ComplementaryApproaches}
To stress how the three complementary approaches can be used in a joint fashion, we will explain the workflow of our study in the following.

An efficient starting point for a study is the unit-cell mean-field approach.
This technique allows to deduce a quantum phase diagram on the mean-field level.
For the LRDIM under investigation, we expect that there are no other phases present in the resulting phase diagram than the ones determined from the mean-field calculations.
Even the order of phase transitions between potential phases can be assessed with this method, by looking for discontinuities in the average magnetization.
Also the mean-field energies in the normal phases are exact due to the absence of quantum fluctuations.
A downside of the approach is the treatment of quantum fluctuations on a mean-field level.
This means that in the phase diagram the extent of the superradiant phases is not quantitative.

To gain further insight into the melting of normal phases, the mapping to the Dicke model is a very capable tool to estimate an upper bound in $g$ for a phase transition.
Further, if the transition point determined from the mapping coincides with the quantitative phase transition point determined by quantum Monte Carlo, one can conclude that the transition is of second order Dicke criticality.
In order to set up the approach described in Sec.~\ref{sec:AlmostEverything}, it is necessary to have insight into the ordering pattern at $g=0$ to perform the respective sublattice rotations (see Eq.~\eqref{eq:LRDIMRotated}).
Since the unit-cell mean-field approach is exact at $g=0$  and only limited by the amount of trial unit cells and the local optimization routines, we use these results to perform the appropriate sublattice rotations.

Regarding the generalized wormhole quantum Monte Carlo, this approach should provide unbiased fully quantitative insight into the entire quantum phase diagram.
An algorithmic inconvenience is that there are no quantum fluctuations in the normal phases.
This leads to an insufficient exploration of the state space at low temperatures and makes the algorithm freeze into metastable normal configurations.
Therefore, it is reasonable to initialize the simulation in this regime in a correct lattice configuration.
As before for the mapping to the Dicke model, these configurations can be easily determined from the unit-cell based mean-field approach which would be computationally too expensive to do with QMC.
Note, that this is a great example where the quantitative quantum Monte Carlo approach, which is biased in the choice of the computational cluster, profits from the qualitative mean-field which is inherently rigorous in the choice of unit-cells.
From the observables determined by the generalized wormhole quantum Monte Carlo it is possible to classify the phase transition by means of finite-size scaling independently from the other approaches.
But, if the phase transition is of second-order Dicke universality, all three methods predict the same quantum critical point.

As a concluding note, this work also functions as a hand-on comparison of the methods available in the toolbox to study models of the LRDIM type.

\section{Results}
In this section we present our results for the ground-state phase diagram of the LRDIM on the square and triangular lattice that were obtained by using the three complementary approaches described in Sec.~\ref{sec:methods}.
The main emphasis lies on the identification of the plethora of emerging ground states, the classification of phase transitions melting the devil's staircase, and the comprehensive calculation of quantitative points of quantum phase transitions.

We studied the model on the square lattice and the triangular lattice for $\alpha=3$ (dipolar interactions) and $\alpha=6$ (van-der-Waals interactions).
We chose to study the LRDIM on the square and triangular lattice as paradigmatic examples where the nearest-neighbor interaction graph is bipartite and non-bipartite, respectively.

\subsection{Square lattice quantum phase diagram}
\label{sec:ResultsSquare}

\newcommand{\LargeCaption}[2]{Quantum phase diagram of the LRDIM on the #1 lattice with exponent $\alpha=#2$. The colors shaded in the background depict the phase diagram determined by the unit-cell based mean-field calculations. The exemplary configurations for the phases in the grid at the bottom of the figure are depicted using the mean-field results at the position of the respective label. The labels in the phase diagram are centered around the point from which the configuration below is taken. The phase boundaries estimated from the mapping to the pure Dicke model are depicted with the black solid lines. The results from the generalized wormhole QMC approach are visualized using the two-colored balls with a symbol in the middle. The two colors indicate the phases at each side of the transition in accordance to colors of the mean-field phase diagram. Below the phase diagram the magnetization profile for $g=0$ is depicted.}

\begin{figure*}[p]
	\centering
	\includegraphics[width=\textwidth]{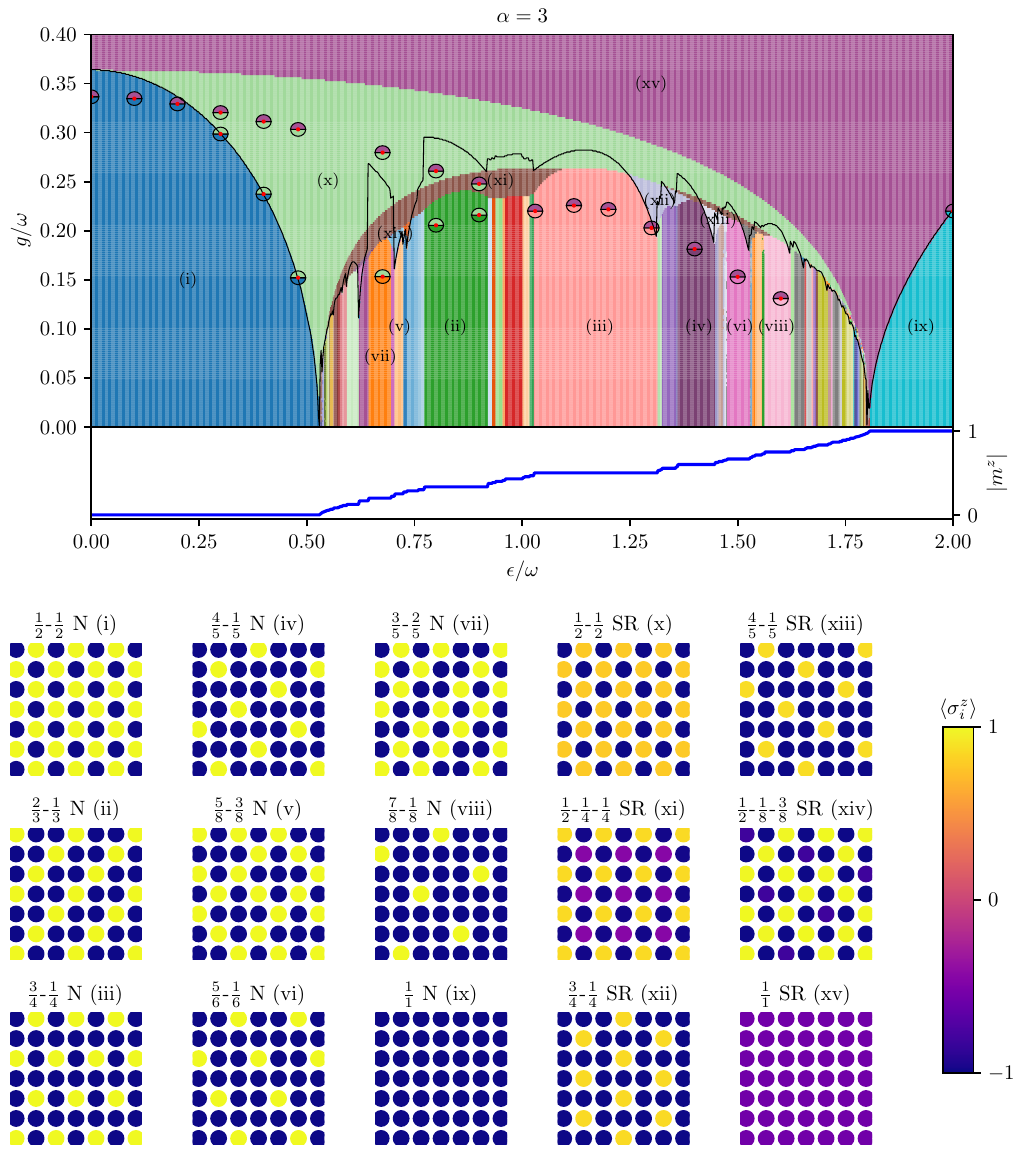}
	\caption{\LargeCaption{square}{3}}
	\label{fig:PDSquareThree}
\end{figure*}

\begin{figure*}[p]
	\centering
	\includegraphics[width=\textwidth]{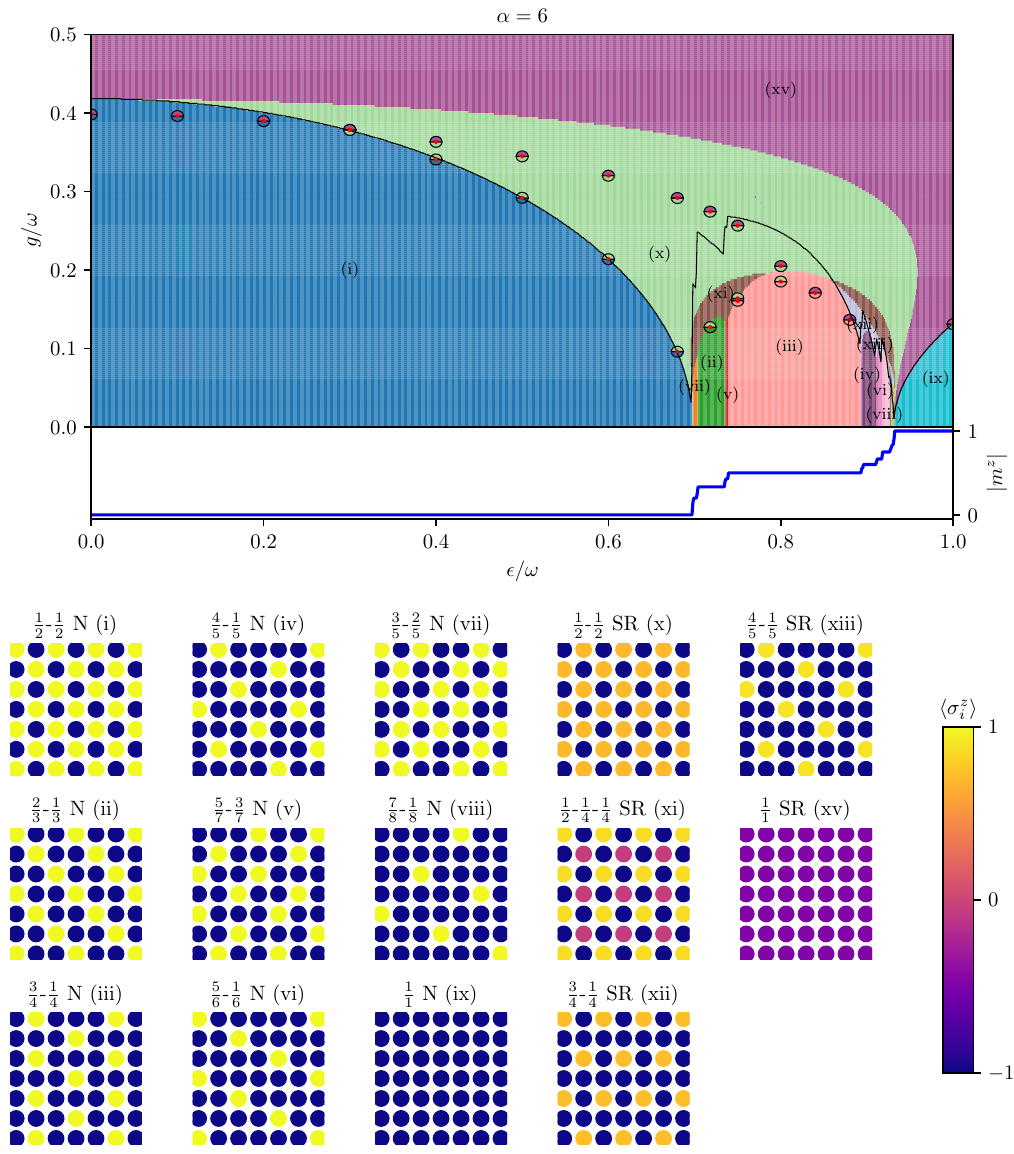}
	\caption{\LargeCaption{square}{6}}
	\label{fig:PDSquareSix}
\end{figure*}

\begin{figure*}[p]
	\centering
	\includegraphics[width=\textwidth]{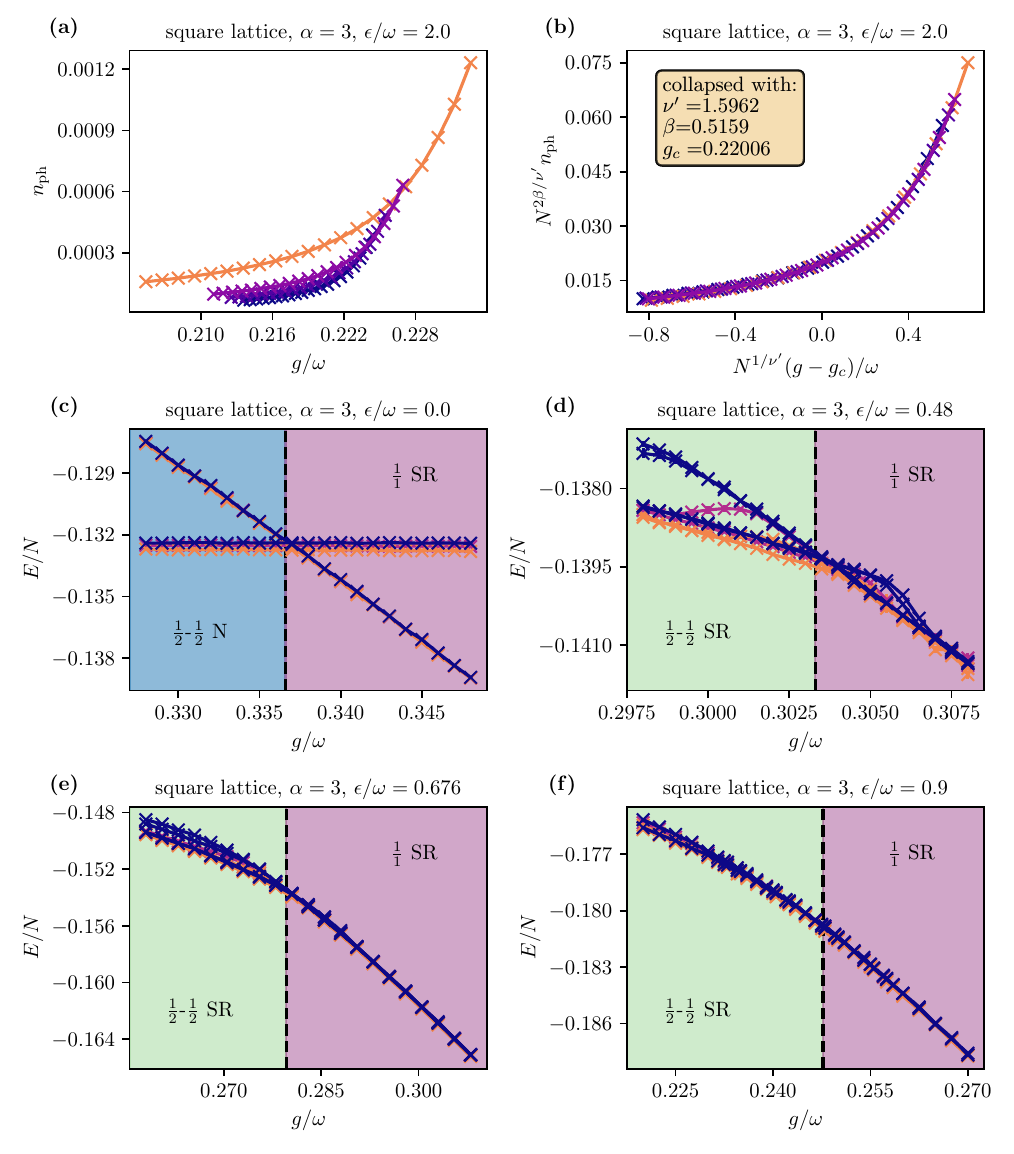}
	\caption{Excerpts of QMC data for a selected subset of the simulated parameters to demonstrate the method for data analysis and softening of first-order transitions. In panel (a) the photon density close to a critical point with Dicke universality is plotted for system size $L\in\{24,36,42,48\}$, while (b) shows the same data rescaled according to finite-size scaling to get a data collapse. The panels (c)-(f) show the energy density close to a phase transition for $L\in\{18,24,36\}$. As the longitudinal field $\epsilon$ is increased, the first-order crossing of the energies of the ground-state phases gets softened. While for (c)-(e) the crossing/kink in the ground-state energy is visible by eye, for $(f)$ one sees effects reminiscent of an avoided level crossing bending the energy density in the left phase $\frac{1}{2}$-$\frac{1}{2}$ prior to the level crossing and it is not clear which type of transition will happen in the thermodynamic limit.}
	\label{fig:PlotQMC}
\end{figure*}

The phase diagrams of the LRDIM on the square lattice for $\alpha=3$ and $\alpha=6$ are shown in Figs.~\ref{fig:PDSquareThree}~and~\ref{fig:PDSquareSix}.
We structure the discussion of the results as follows:
First, we graze over the limiting cases.
Second, we indulge into the intermediate regime and the details on how the devil's staircase at $g=0$ breaks down towards superradiant phases.

When regarding Figs.~\ref{fig:PDSquareThree}~and~\ref{fig:PDSquareSix} one sees the devil's staircase of magnetization plateaux at $g=0$, which is also depicted as a corresponding plot of magnetization against longitudinal field directly below the phase diagram.
In general, the staircase for $\alpha=3$ has a larger extent in $\epsilon$ due to the larger effective coordination number \cite{Koziol2023,Koziol2024}.
By the symmetry of the Hamiltonian, the phase diagram for $\epsilon>0$ is identical to the one for $\epsilon<0$ with $\sigma_i^z\rightarrow -\sigma_i^z$.
We see that in the devil's staircase the $\frac{1}{2}$-$\frac{1}{2}$ N phase is the most prominent, because it is stabilized by the nearest-neighbor Ising interaction.
We observe that the proportion of this phase in the staircase diminishes with decreasing $\alpha$.
For $\alpha=6$, other dominant plateaus are the $\frac{2}{3}$-$\frac{1}{3}$ N and $\frac{3}{4}$-$\frac{1}{4}$ N, as well as, the smaller plateaux  $\frac{4}{5}$-$\frac{1}{5}$ N,  $\frac{5}{6}$-$\frac{1}{6}$ N, and  $\frac{7}{8}$-$\frac{1}{8}$ N.
These plateaus are also present for $\alpha=3$, but a plethora of others with a similar extent appear.
For sufficiently strong magnetic field $\epsilon$, the magnetization saturates to $\abs{m^z}=1$ in a completely field-polarized normal phase.

In the limit of large $g$, a superradiant phase occurs without any magnetic structure in the matter degrees of freedom.
As expected, we find a second-order Dicke phase transition between the field-polarized normal phase and this superradiant phase independent of the $\alpha$ value.
We see that the estimates for the critical point obtained by all three methods are in agreement.
Following the mapping to the Dicke model, this means that the hybridized excitations of the $\vec k = (0,0)$ magnon mode and the photonic cavity mode \cite{Schellenberger2024} condensate at the quantum critical point driving the Dicke phase transition.
Further, we confirm the Dicke phase transition from the Monte Carlo data, using data collapses of the photon density \cite{Langheld2022,Adelhardt2024}.
An exemplary data collapse for the transition between the fully field-polarized $\frac{1}{1}$ N and $\frac{1}{1}$ SR phase is shown in Fig.~\hyperref[fig:PlotQMC]{\ref*{fig:PlotQMC}a)-b)} for $\alpha=3$ and $\epsilon = 2\omega$. 
In Fig.~\hyperref[fig:PlotQMC]{\ref*{fig:PlotQMC}a)} we show the unscaled curves used for the data collapse and in Fig.~\hyperref[fig:PlotQMC]{\ref*{fig:PlotQMC}b)} we display the same curves rescaled with the fitted exponents.
Note, the Dicke transition can be interpreted as a quantum phase transition above the upper critical dimension \cite{Langheld2024}, therefore special finite-size scaling techniques for this class of quantum phase transitions are required \cite{Kenna2013,Koziol2021,Langheld2022,Berche2022,Adelhardt2024}. 
We choose the consistent definition of critical exponents $\beta$ and $\nu^\prime$ from Ref.~\cite{Langheld2024}.

Now, to discuss the melting of the devil's staircase we first discuss the qualitative predictions of the unit-cell based mean-field approach to provide a general overview over potential scenarios occurring.
In a second step, we will challenge these predictions with the fully quantitative quantum Monte Carlo results to see which phases withstand the full incorporation of quantum fluctuations.

In the unit-cell based mean-field calculations, the entire devil's staircase is enveloped by the $\frac{1}{2}$-$\frac{1}{2}$ SR phase. 
This phase has the same symmetry in the matter degrees of freedom as the $\frac{1}{2}$-$\frac{1}{2}$ N phase.
Since a transition between these phases requires the condensation of the bosons into the ground state, we expect and confirm the transition to be in the Dicke universality class.
Further, we find the gapped phases at $g=0$ in the plateaux of the staircase to have a finite extent for $g>0$.
We observe that on the mean-field level, the plateaux with a magnetization $0 < \abs{m^z} < 1$ are to a large extent enveloped by a $\frac{1}{2}$-$\frac{1}{4}$-$\frac{1}{4}$ SR phase having an ordering pattern with three sublattices.
Additionally, there are several phases, e.\,g., the $\frac{3}{4}$-$\frac{1}{4}$ N, $\frac{4}{5}$-$\frac{1}{5}$ N or $\frac{5}{6}$-$\frac{1}{6}$ N, that have an additional supperradiant phase at the tip of the lobe with the same symmetry in the matter degrees of freedom.

On the one hand, we see that the quantum phase transitions between normal and superradiant phases with the same spatial symmetry have a second-order Dicke phase transition between them. 
On the other hand, the melting transition into phases with non-matching symmetry is of first order. 
This behavior is already visible in the mean-field calculations, but also confirmed by the mapping to the effective Dicke model and the quantitative QMC calculations.

The main point in the following will be to assess with the quantum Monte Carlo approach which of the intermediate superradiant phases predicted by the mean-field approach survive when treating the quantum fluctuations quantitatively. 
Beyond the intermediate $\frac{1}{2}$-$\frac{1}{2}$ SR phase, we will have an emphasis on the possibility of the $\frac{3}{4}$-$\frac{1}{4}$ SR phase and the $\frac{1}{2}$-$\frac{1}{4}$-$\frac{1}{4}$ SR phase since they have the largest region of stability on the mean-field level.
One general trend one can extract from the QMC data is that the magnetically uniform superradiant phase has a larger extent than predicted by the mean-field analysis and significantly reduces the extent of intermediate phases.
The envelope of the normal phases by the $\frac{1}{2}$-$\frac{1}{2}$ SR phase is severely reduced for $\epsilon \leq 0.9\omega$ ($\epsilon \leq 0.8\omega$) and seems to be completely absent for $\epsilon \geq 1.03\omega$ ($\epsilon \geq 0.84\omega$) for $\alpha=3$ ($\alpha=6$).
Also the tip of the $\frac{1}{2}$-$\frac{1}{2}$ N phase is cut off by the $\frac{1}{1}$ SR phase and the first-order transition between these two phases at $\epsilon=0$ extends to finite $\epsilon \leq 0.2\omega$ ($\epsilon \leq 0.2\omega$) for $\alpha=3$ ($\alpha = 6$).

Finding traces of intermediate superradiant phases with more complex ordering patterns than a checkerboard structure is harder as these phases are already a lot smaller in the mean-field calculation than the $\frac{1}{2}$-$\frac{1}{2}$ SR phase.
The QMC data cannot exclude the existence of a $\frac{3}{4}$-$\frac{1}{4}$ SR phase on the right side of the $\frac{3}{4}$-$\frac{1}{4}$ N phase, but can restrict it to a very small notch on the right side of the $\frac{3}{4}$-$\frac{1}{4}$ N lobe.
Resolving this is difficult and computationally expensive. Not only because the extent of the phase is tiny but also because one needs large inverse temperature $\beta$ due to the closeness of other states of the devil's staircase.
For $\alpha=6$, the existence of the $\frac{1}{2}$-$\frac{1}{4}$-$\frac{1}{4}$ SR phase was found in a tiny region for $\epsilon = 0.75\omega$ for $ 0.1606(3) \lessapprox g \lessapprox 0.1637(3)$.
We have also checked the stability of this phase for $\epsilon/\omega = 0.676, 0.80, 0.90$ ($\epsilon/\omega = 0.718, 0.8$) for $\alpha=3$ ($\alpha=6$) close to the transitions between the normal phases of the devil's staircase and the $\frac{1}{2}$-$\frac{1}{2}$ SR phase. 
For these parameters we found no indication of this state being the ground state although the energy oftentimes appeared to be close to the ones of the respective ground states.

Another special point is the melting from the $\frac{3}{5}$-$\frac{2}{5}$ N phase for $\alpha=3$ (vii), where a complex $\frac{1}{2}$-$\frac{3}{8}$-$\frac{1}{8}$ SR phase (xiv) or even $\frac{1}{2}$-$\frac{1}{4}$-$\frac{1}{8}$-$\frac{1}{8}$ SR phase lies close according to mean field.  
In the correlation functions obtained by QMC, we see fluctuations locally forming such patterns close to the melting point of the $\frac{3}{5}$-$\frac{2}{5}$ N phase.
The lowest energy was achieved by a $\frac{1}{2}$-$\frac{1}{2}$ SR phase, but the QMC algorithm has trouble equilibrating in this parameter regime due to a lack of fluctuations and we cannot rule out that a more complex phase could in fact be favorable. We therefore include the crossing point of the energies of $\frac{3}{5}$-$\frac{2}{5}$ N and $\frac{1}{2}$-$\frac{1}{2}$ SR phase in the phase diagram but it should be taken with a grain of salt.

The transition from the lobes to the $\frac{1}{1}$ SR phase with no magnetic ordering is commonly found to be a first-order transition.
In particular if the transition happens between a normal phase with magnetic ordering and the superradiant phase without magnetic ordering, the transition has to be of first order as the adjacent phases break different symmetries of the Hamiltonian.
The transition between the intermediate $\frac{1}{2}$-$\frac{1}{2}$ SR phase and the $\frac{1}{1}$ SR phase appears to be of first order for smaller values of $\epsilon$ but the level crossing gradually softens for increasing $\epsilon$.
For $\epsilon \geq 0.8\omega$ ($\epsilon \geq 0.75\omega$) for $\alpha = 3$ ($\alpha=6$), the crossing of the ground-state energies is so flat that it becomes hard to distinguish it from a second-order transition. 
This softening of the crossing in the energy density is depicted in Fig.~\hyperref[fig:PlotQMC]{\ref*{fig:PlotQMC}c)-f)} for $\alpha = 3$ for a selection of $\epsilon$ values. 
While at $\epsilon = 0$ (Fig.~\hyperref[fig:PlotQMC]{\ref*{fig:PlotQMC}c)}) the crossing is very sharp, the transition becomes less sharp for $\epsilon/\omega = 0.48, 0.676$ (Fig.~\hyperref[fig:PlotQMC]{\ref*{fig:PlotQMC}d)-e)}) but still exhibits the kink of a first-order transition. 
For $\epsilon/\omega = 0.9$ (Fig.~\hyperref[fig:PlotQMC]{\ref*{fig:PlotQMC}f)}) the energy density experiences rounding on the left of the phase transition and it is hard to tell whether the transition will be of first or second order in the thermodynamic limit.
For small system sizes at $\epsilon=0.9\omega$, several observables like the magnetization, Binder cumulant and energy suggest a second-order transition with 3D Ising criticality, while the larger system sizes display weak hystereses.
It could be that the condensation of elementary excitations and a level crossing are close to each other.
In this case, it is to be expected that the smaller systems are more affected by the finite-size rounding close to the avoided level crossing and therefore show characteristics of a second-order transition, while the large systems are not as affected by the rounding. 
However, we cannot rule out that the algorithm has difficulties with large system sizes at these exceptional points and the small systems actually display the correct behavior of the thermodynamic limit.

As explained in Sec.~\ref{sec:MonteCarlo}, we use a swipe mechanism starting in one phase with a spin configuration close to the expected ground state and tuning the photon coupling $g$ across the phase boundary. 
This is necessary for most of the first-order transitions as one needs a global change in the spin configuration for which the local optimal energy might not correspond to the global minimum for the whole system.
For second-order transitions, this is not necessary as the system-wide fluctuations close to the phase transition drive the system from one phase to the other.
However, even though the algorithm has trouble converging to the perfect spin pattern globally, one can still get an idea of the optimal ordering pattern of the spins by looking at the correlation function and identifying repetitive patterns that are sheared/rotated against each other.

As predicted by the mapping of non-SR phases to the Dicke model, the transitions from non-SR phases to the SR phase with the same symmetry-breaking in the matter is of second order and has the same criticality as the Dicke model. 
In some parameter regions, the data collapse from the QMC data has a bias towards larger $\nu'$ (and therefore also large $\beta$ as they are correlated, while the fraction $\beta/\nu'$ does not suffer from such bias), which determines the finite-size rounding in the tuning parameter.  The affected transitions have in common that they were either very close to another transition or at low $g_c$.
For less problematic regions like the large-$\epsilon$ case with a transition from a $\frac{1}{1}$ N to $\frac{1}{1}$ SR phase, we are able to derive the correct exponents up to an adequate accuracy with $\nu^\prime_6(\epsilon=1)= 1.64(9)$ and $\beta_6(\epsilon=1)=0.49(4)$ and $\nu^\prime_3(\epsilon = 2) = 1.596(18)$ and $\beta_3(\epsilon = 2) = 0.516(13)$ in comparison to the expected mean-field values of $\nu'=3/2$ and $\beta=1/2$ \cite{Langheld2024}.

Overall, we conclude that the predictions from the mean-field analysis turned out to be qualitatively accurate. 
Even though many of the intermediate phases could not be found with QMC and are probably eradicated by the prominent $\frac{1}{1}$ SR or $\frac{1}{2}$-$\frac{1}{2}$ SR phases, the discovered intermediate phases were found close to where mean-field predicted them to appear.
Moreover, the predictions from the mean-field calculation aided the other methods. In particular the exact ordering structure at $g=0$ was an indispensable ingredient for the mapping of non-SR phases and the QMC approach.
While all three methods had their strengths and weaknesses, they complemented each other in a nice fashion, e.\,g., mean field being exact at $g=0$ where QMC is not competitive due to the lack of fluctuations and, in contrast, QMC performing better in the SR phases due to the fluctuations introduced by the bosons where mean field is not exact anymore.

\subsection{Triangular lattice quantum phase diagram}
\label{sec:ResultsTriangular}

\begin{figure*}[p]
	\centering
	\includegraphics[width=\textwidth]{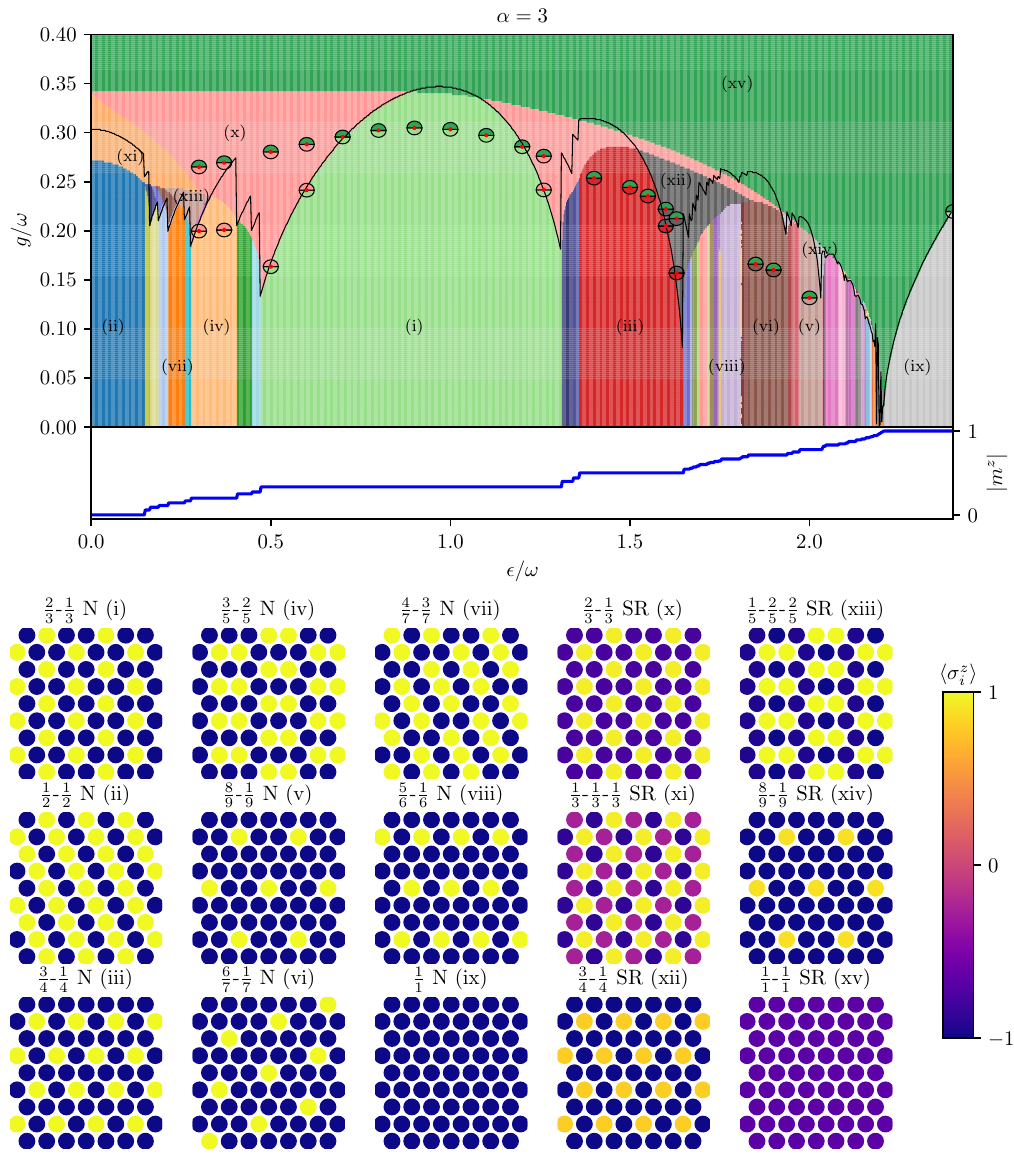}
	\caption{\LargeCaption{triangular}{3}}
	\label{fig:TriangularThree}
\end{figure*}

\begin{figure*}[p]
	\centering
	\includegraphics[width=\textwidth]{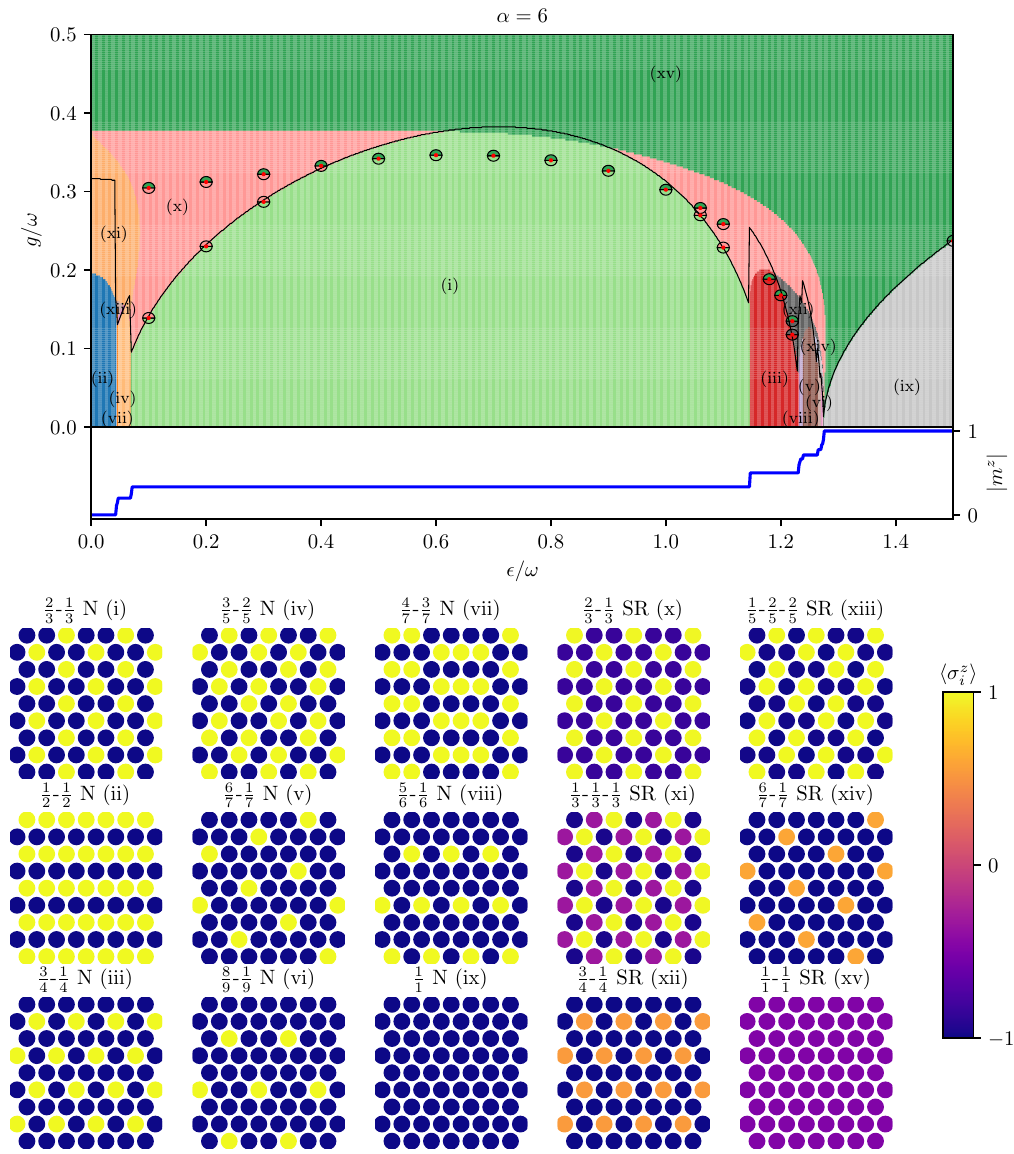}
	\caption{\LargeCaption{triangular}{6}}
	\label{fig:TriangularSix}
\end{figure*}

The quantum phase diagrams of the LRDIM on the triangular lattice for $\alpha=3$ and $\alpha=6$ are presented in Figs.~\ref{fig:TriangularThree}~and~\ref{fig:TriangularSix}.
As the discussion has many similarities to the square lattice discussed in Sec.~\ref{sec:ResultsSquare}, we will focus on the occurring differences between the two lattices.

For $g=0$, we see the devil's staircase of magnetization plateaus, similar to the square lattice.
The main difference is that the $\frac{2}{3}$-$\frac{1}{3}$ N phase is the dominant phase, because in the nearest-neighbor limit only this phase remains from the staircase.
As for the square lattice, the extent of the staircase in $\epsilon$ increases with decreasing $\alpha$ value and the proportion of the plateaux besides the $\frac{2}{3}$-$\frac{1}{3}$ phase increases.

Equivalent to the square lattice, there is a field-polarized normal phase for sufficiently large $\epsilon$ values which melts via a Dicke transition to the uniform superradiant phase.
Furthermore, at sufficiently large $g$ values, the system is in a superradiant phase with no order in the matter degrees of freedom.

Regarding the melting of the devil's staircase on the mean-field level, we observe the following:
The majority of the devil's staircase is enveloped by the $\frac{2}{3}$-$\frac{1}{3}$ SR phase, besides a small but finite window around the tip of the $\frac{2}{3}$-$\frac{1}{3}$ N phase where we observe a direct first-order transition to the uniform SR phase.
Further remarkable plateaux of the devil's staircase are the hexagonal Wigner crystal phases $\frac{3}{4}$-$\frac{1}{4}$ N, $\frac{6}{7}$-$\frac{1}{7}$ N, and $\frac{8}{9}$-$\frac{1}{9}$ N, as well as, the $\frac{1}{2}$-$\frac{1}{2}$ N plain stripes around $\epsilon=0$ and the other structures depicted in Figs.~\ref{fig:TriangularThree}~and~\ref{fig:TriangularSix}.
Note, the $\frac{1}{2}$-$\frac{1}{2}$ N phase is not present in the nearest-neighbor limit of the model due to the frustration in this limit.
It is established that long-range interactions stabilize the formation of plain stripes \cite{Metcalf1974,Kaburagi1974,Korshunov2005,Smerald2016,Smerald2018,Koziol2019,Koziol2021,Koziol2023}.
Further, analogous to the XXZ model with long-range Ising interactions \cite{Koziol2024}, the plain stripes undergo a transition to a three-sublattice ordered phase on the mean-field level.
Here, this phase is a $\frac{1}{3}$-$\frac{1}{3}$-$\frac{1}{3}$ SR phase.
On the mean-field level the plateaux determined between the $\frac{1}{2}$-$\frac{1}{2}$ N, and the $\frac{3}{5}$-$\frac{2}{5}$ N phase have second-order phase transition to SR phases with the same magnetic order on the left flanks of the lobes (see Figs.~\ref{fig:TriangularThree}~and~\ref{fig:TriangularSix}). 
In general, the mean-field phase diagram of intermediate phases between the $\frac{1}{2}$-$\frac{1}{2}$ N, the $\frac{3}{5}$-$\frac{2}{5}$ N and the uniform superradiant phase has strong resemblance to the mean-field phase diagram of the XXZ magnet with long-range Ising interactions presented in Ref.~\cite{Koziol2024}.
In the long-range Dicke-Ising model the intermediate phases are magnetically structured superradiant phases while in Ref.~\cite{Koziol2024} these phases are supersolid phases.

As for the square lattice, there are two scenarios how a normal phase of the staircase melts towards a superradiant phase: First, via a Dicke transition to a superradiant phase with matching order. Second, via a first-order transition.

Notably, on the flank towards larger $\epsilon$-values, the $\frac{3}{4}$-$\frac{1}{4}$ N phase has a second-order Dicke transition to the $\frac{3}{4}$-$\frac{1}{4}$ SR phase on the mean-field level.

In the following, we will discuss the quantitative QMC results shown in Figs.~\ref{fig:TriangularThree}~and~\ref{fig:TriangularSix}.
Similar to the results for the square lattice, the QMC simulations show the general trend that the uniform superradiant phase has a larger extent than predicted by the mean-field analysis.
In the figures we track the transition line from this SR phase to the low-field phases and report that the envelope by the $\frac{2}{3}$-$\frac{1}{3}$ SR is reduced at the left and severely reduced at the right flank of the $\frac{2}{3}$-$\frac{1}{3}$ N phase. 
It is completely absent for high $\epsilon$ values in the devil's staircase starting with the $\frac{3}{4}$-$\frac{1}{4}$ N phase.
Further, the tip of the $\frac{2}{3}$-$\frac{1}{3}$ N, where a direct first-order transition to the uniform SR phase takes place, is of larger extent in $\epsilon$ compared to mean field.

The first-order character of the envelope transition line to the uniform $\frac{1}{1}$ SR phase persists also beyond the $\frac{2}{3}$-$\frac{1}{3}$ normal lobe, i.\,e.\,, for the transition between the intermediate $\frac{2}{3}$-$\frac{1}{3}$ SR and uniform $\frac{1}{1}$ SR phase. 
Similar to the square lattice, there is also a softening of the level crossing on the left side of the lobe which gets stronger for smaller $\epsilon$.
However, the softening in the case of the triangular lattice is not as severe as for the square lattice and all investigated transitions still exhibit a visible kink in the ground-state energy.

We verify with the QMC calculations that the phase transition between the $\frac{2}{3}$-$\frac{1}{3}$ N and $\frac{2}{3}$-$\frac{1}{3}$ SR phase is a second-order Dicke transition, although the critical exponents extracted from the data collapse are not as good as for the transition between uniform N and SR phases. 
Especially for transitions at low $g$ on the edge of lobes, the extracted critical exponents for the transition become worse as it was already the case for the square lattice.

We would like to mention that we deliberately did not apply the QMC approach for $\epsilon< 0.3\omega$ ($\epsilon < 0.1\omega$) for $\alpha=3$ ($\alpha=6$).
In this regime, close to zero longitudinal field, the algorithm struggles from issues known from the study of the related antiferromagnetic long-range transverse-field Ising models (TFIM) on the triangular lattice: The three sublattice structure $\frac{1}{3}$-$\frac{1}{3}$-$\frac{1}{3}$ SR is related to the $\sqrt{3}\times\sqrt{3}$-clock order in the TFIM, which is not easily captured by algorithms relying on "bond-decompositions" \cite{Humeniuk2016,Biswas2016,Biswas2018}. 
We excluded this regime here as it would deserve a study on its own, maybe even adapting the algorithm accordingly which is beyond the scope of this paper.
For the nearest-neighbor TFIM there are ideas how to enhance the capabilities of QMC approaches by decompositions of the Hamiltonian into physically more relevant motifs \cite{Biswas2016,Biswas2018}.
But even for the long-range TFIM, a large-scale QMC study resolving the plain stripes and a three-sublattice order has not been conducted.
For the Dicke-Ising model even the behavior in the nearest-neighbor case in this regime is yet to be studied.
Therefore, we exclude this regime from the quantitative analysis in this work.

Similar to the square lattice, one purpose of the QMC approach is to verify if intermediate superradiant phases with more complex ordering patterns in the matter degrees of freedom occur in the true quantum phase diagram.
This task is especially hard since even in the mean-field analysis several of these phases already have a very small extent.
We verify and show in Figs.~\ref{fig:TriangularThree}~and~\ref{fig:TriangularSix} that there is an intermediate $\frac{3}{4}$-$\frac{1}{4}$ SR phase on the lower part of the right flank of the $\frac{3}{4}$-$\frac{1}{4}$ N phase for $\alpha=3$ and $\alpha=6$.
One can understand this phase as a ``superradiant Wigner crystal'' with a unit cell of four sites.
We report a second-order Dicke transition from the $\frac{3}{4}$-$\frac{1}{4}$ N phase to the $\frac{3}{4}$-$\frac{1}{4}$ SR phase as expected from the mapping of non-SR phases to the Dicke model and the mean-field analysis.
At larger $g$ values, the QMC shows that the $\frac{3}{4}$-$\frac{1}{4}$ SR phase has a first-order transition directly to the uniform superradiant phase.
In contrast, the mean-field calculations still see a $\frac{2}{3}$-$\frac{1}{3}$ SR phase inbetween.
We have specifically studied this scenario in the QMC approach and found no remainders of this phase for any studied transition between $\frac{3}{4}$-$\frac{1}{4}$ N or SR phases and the uniform superradiant phase.
We were not able to resolve any further intermediate phases on the triangular lattice, albeit the interesting -- low $\epsilon$ -- regime is not yet within reach.

To wrap up, we conclude for the LRDIM on the triangular lattice that overall the predictions from the mean-field analysis turned out to be qualitatively accurate.
Even though many of the predicted intermediate phases could not be found in the QMC simulations, we showed that the $\frac{3}{4}$-$\frac{1}{4}$ SR phase is stabilized by the long-range interactions.
In the low $\epsilon$ regime, the unit-cell-based mean-field approach provides a first glance at possible intermediate phases which are yet to be verified by quantitative methods.

\section{Summary and outlook}
\label{sec:summaryandoutlook}

In this work we determined the quantitative ground-state phase diagram of the Dicke Ising model with algebraically decaying antiferromagnetic long-range interactions ($\alpha=3$ and $\alpha=6$) on the square and triangular lattice.
To study the model we developed and employed a trident of complementary tools: a cluster based mean-field approach, an exact mapping of normal phases to the Dicke model, and a generalized wormhole quantum Monte Carlo approach.

In the limit of a vanishing light-matter coupling we reproduce the devil's staircase of magnetically ordered phases as expected from a long-range Ising model in a longitudinal field.

The focus of this work lies on studying the breakdown of the devil's staircase of magnetically ordered normal phases towards superradiant phases.
Here, we observe two distinct scenarios how normal phases melt towards a superradiant phase with respect to a linear coupling to a single bosonic mode:
First, if the normal and superradiant phase have the same underlying magnetic order, the transition will be of Dicke universality and all three approaches employed in this work result in the same critical point.
This phase transition is driven by the boson-dressed magnon condensation into the ground state.
Second, if the normal and superradiant phase have a different underlying magnetic order, the transitions are first-order level-crossing transitions.
Further, magnetically ordered superradiant phases will eventually transition to the uniform superradiant phase.
Transitions between superradiant phases are generically first-order transitions, although we report indications of a change in the order of the transition line to the uniform superradiant phase towards a second-order transitions with 3D Ising criticality.

We further highlight the appearance of intermediate superradiant phases with exotic magnetic orders which stabilize due to the long-range interactions.
One can see these phases as ``superradiant Wigner crystals''.
Notable examples from our study are the $\frac{3}{4}$-$\frac{1}{4}$ SR phase on the triangular lattice and the three sublattice $\frac{1}{2}$-$\frac{1}{4}$-$\frac{1}{4}$ SR phase on the square lattice.
Since the mean-field study demonstrates the possibility of further SR phases, a possible direction for future research is to tailor interactions in a specific way to stabilize desired superradiant Wigner crystals.

Regarding the experimental implementation of the long-range Dicke Ising model, we stress the existence of no-go theorems for many proposed equilibrium setups \cite{Rzazewski1975,Bialynicki1979,Gawedzki1981,Rzazewski1991,Keeling2007,Nataf2010A,Nataf2010B,Vukics2012,Viehmann2011,Bamba2016,Andolina2019,Andolina2022}. 
Therefore, we suggest that the model should be realized as an effective description of non-equilibrium systems \cite{Dimer2007,Baumann2010,Bastidas2012,Gelhausen2016,Zhiqiang2017,Zhang2018,Puel2024} circumventing these issues.
Since long-range van-der-Waals interactions are generically present in Rydberg atom quantum simulators \cite{Browaeys2020} we suggest the exploration of Rydberg-dressed spin lattices coupled to a single cavity mode \cite{Gelhausen2016,Puel2024} via cavity-assisted Raman transitions \cite{Dimer2007,Zhiqiang2017,Zhang2018} as a realization platform.

\section{Acknowledgements}

We thank Giovanna Morigi, Andreas Schellenberger, and Tom Schmidt for fruitful discussions.
We further thank Andreas Schellenberger for his critical reading of the manuscript and numerous valuable suggestions to improve it.
We greatfully acknowledge the scientific support and HPC resources provided by the Erlangen National High Performance Computing Center (NHR@FAU) of the Friedrich-Alexander-Universität Erlangen-Nürnberg (FAU). 

\subsection*{Funding information}
This work was funded by the Deutsche Forschungsgemeinschaft (DFG, German Research Foundation) - Project-ID 429529648 - TRR 306 QuCoLiMa (Quantum Cooperativity of Light and Matter). We acknowledge the support by the Munich Quantum Valley, which is supported by the Bavarian state government with funds from the Hightech Agenda Bayern Plus. The hardware of NHR@FAU is funded by the German Research Foundation DFG. 

\subsection*{Author contribution}
All authors contributed to the conceptualization of the study.
JAK developed and implemented the unit-cell-based mean-field approach and performed the calculations.
JAK implemented the mapping of non-superradiant phases to the Dicke model and performed the calculations.
AL developed and implemented the generalized wormhole quantum Monte Carlo for long-range interacting Ising interactions and performed the calculations.
JAK and AL wrote the original draft and visualized the data.
KPS supervised the work.
All authors contributed to the interpretation of the results and the review and editing of the manuscript.

\bibliographystyle{apsrev4-2}
\bibliography{bibliography}

\end{document}